\documentclass[aps,prd,showpacs,floatfix,preprint]{revtex4}
\usepackage{amsmath,bm}
\usepackage{graphicx}
\usepackage{epsfig}
\begin{document}

\title{Isospin dependence of entrainment in superfluid neutron stars in a 
relativistic model}

\author{Apurba Kheto and Debades Bandyopadhyay}
\affiliation{Astroparticle Physics and Cosmology Division and Centre for 
Astroparticle Physics, Saha Institute of Nuclear Physics, 1/AF Bidhannagar, 
Kolkata-700064, India}

\begin{abstract}
We study the entrainment effect between superfluid neutrons and charge neutral 
fluid (called the proton fluid) which is made of protons and electrons in
neutron star interior within the two fluid formalism and using a relativistic 
model where baryon-baryon interaction 
is mediated by the exchange of $\sigma$, $\omega$ and $\rho$ mesons. This model
of strong interaction also includes scalar self interactions. The entrainment 
matrix and entrainment parameter are calculated using the parameter sets of 
Glendenning (GL) and another non-linear (NL3) interaction. The inclusion of 
$\rho$ mesons strongly 
influences the 
entrainment parameter ($\epsilon_{mom}$) in a superfluid neutron star. The 
entrainment parameter is constant at the core and drops rapidly at the 
surface. It takes values within the physical range.    
\pacs{97.60.Jd, 26.60.+c, 47.75.+f, 95.30.Sf}
\end{abstract}

\maketitle

\section{Introduction}

Novel phases of dense matter might exist in neutron stars. Superfluidity or
superconductivity is one such form of matter. Recently, rapid cooling
of the neutron star in Cassiopeia A was reported \cite{ho}. This has been 
attributed to the neutron superfluidity in neutron star cores \cite{page}. 
Pulsar glitches are also thought to be the manifestation of superfluidity in
the crust and core of neutron stars \cite{baym,itoh,and,cha}. This glitch
phenomenon might be described based on the pinning and unpinning of superfluid 
quantized vortices in neutron stars. 

Superfluidity in neutron stars was studied in great detail in Newtonian as
well as general relativistic formulations \cite{prix,andc,cha08}. The 
fluid formalism in the case of superfluidity is different from that of the
perfect fluid. For neutron stars made of neutrons, protons and electrons, two
fluid formalism was used to describe the superfluidity in neutron star matter
\cite{andc}. In this case, one fluid is the superfluid neutrons and the other
fluid called the proton fluid represents the charge neutral component made of 
protons and electrons. It is a well known fact that two fluids in a mixture 
are not decoupled when one fluid interpenetrates through the other. 
In this situation, the momentum of one fluid is proportional to the linear
combination of the velocities of both fluids. This effect is known as 
entrainment.
The entrainment effect has been studied intensively in understanding rotational
equilibria, oscillations of superfluid neutron stars 
\cite{prix,lan02,and01,acp,sota} and the pulsar glitch \cite{and,cha}. 

Superfluid dynamics including entrainment in neutron stars were studied using
Newtonian calculations. The entrainment effect in nonrelativistic and
relativistic Fermi-liquid models 
was studied by different groups \cite{boru,haen,gus05,gus09}. 
A Similar model was developed for entrainment to study slowly
rotating superfluid Newtonian neutron stars \cite{prix}. Comer and Joynt 
calculated the entrainment effect in a relativistic field theoretical model
and obtained first order corrections to the slowly rotating superfluid neutron
stars \cite{joy} for the first time. However, the relativistic model was 
inadequate to describe the neutron star matter because the $\sigma$-$\omega$
Walecka model was adopted in this calculation. Neutron star matter is highly
asymmetric and the inclusion of $\rho$ mesons in the Walecka model is absolutely
necessary. Consequently, it is worth studying the effects of symmetry energy on
the master
function and superfluid dynamics in neutron stars. This motivates us to extend
the calculation of Comer and Joynt \cite{joy,com04} to include 
$\rho$ mesons along with scalar self-interactions.  

We organise the paper in the following way. We describe the relativistic
$\sigma$-$\omega$-$\rho$ model for entrainment and the connection between
the master function and relativistic mean field model as well as the formalism
for slowly rotating superfluid neutron stars in Sec. II.
Results of this calculation are discussed in Sec. III. Section IV gives the
summary and conclusions.

\section{Formalism}

Here we adopt the two fluid formalism as described
in Refs.\cite{com01,com02,car,lan94,lan95,lan982,lan983,lan99} to study 
the entrainment effect in cold neutron stars.
The signature of the metric used here is the same as in Ref.\cite{joy}. 
The starting point of the superfluid formalism is the master function 
($\Lambda$). It is a function of scalars which are constructed from neutron
($n^{\mu}$) and proton ($p^{\mu}$) number density currents such as
$n^2= - n_\mu n^\mu$, $p^2= - p_\mu p^\mu$ and  $x^2= - n_\mu p^\mu$.
The master function $-\Lambda (n^2,p^2,x^2)$ 
corresponds to the total thermodynamic energy density when neutron and proton
currents flow in parallel.
The stress-energy tensor in terms of the master function is written as 
\cite{joy,com04}
\begin{equation}
T^{\mu}_\nu = \Psi {\delta^\mu _\nu} + n^\mu \mu_\nu + p^\mu \chi_\nu
\end{equation}
where
\begin{equation}
    \Psi = \Lambda - n^\rho \mu_\rho - p^\rho \chi_\rho 
\end{equation}
is the generalized pressure, and
\begin{eqnarray}
\label{coef1}
    \mu_\nu &=& {\cal B} n_\nu + {\cal A} p_\nu  \ , \\
    \chi_\nu &=& {\cal A} n_\nu + {\cal C} p_\nu \ , 
\label{coef2}
\end{eqnarray}
are neutron and proton momentum covectors and are conjugate to $n^{\mu}$ and 
$p^{\mu}$, respectively. It implies that neutron or proton momentum is a linear
combination
of both number density currents. The entrainment effect disappears when the 
coefficient ${\cal A}$ is zero. The magnitudes of momentum covectors $\mu_\nu$
and $\chi_\nu$ are chemical potentials of neutron and proton fluids,
respectively \cite{joy}. The proton fluid is a charge neutral fluid made
of protons and electrons \cite{lan99}. For $\beta$-equilibrated neutron star 
matter, chemical potentials of neutron and proton fluids are the same. 
Consequently, the chemical potential of 
proton fluid is the sum of chemical potentials of protons and electrons 
\cite{lan99,rieu}.
Coefficients in Eqs. ({\ref{coef1}}) and
({\ref{coef2}}) are determined from the master function,      
\begin{equation}
    {\cal A} = -\frac {\partial \Lambda} {\partial x^2}~, 
    {\cal B} = -2\frac {\partial \Lambda} {\partial n^2}~, 
    {\cal C} = -2\frac { \partial \Lambda} {\partial p^2}~. 
\end{equation}
The field equations for neutrons and protons involve two conservation and
two Euler equations. 

The  master function is determined from averaged stress-energy components
in a covariant way from the following relation 
\cite{joy,com04}
\begin{equation}
    \Lambda = - {1 \over 2} \left \langle T \right\rangle + 
              {3 \over 2} \left(x^4 - n^2 p^2\right)^{- 1} \left(
              n^2 p^2 \left[{1 \over n^2} n^{\mu} n^{\nu} + 
              {1 \over p^2} p^{\mu} p^{\nu}\right] - x^2 \left[
              n^{\mu} p^{\nu} + p^{\mu} n^{\nu}\right]\right) 
              \left\langle T_{\mu \nu} \right\rangle~,
\label{lcov}
\end{equation}  
where 
$\left \langle T \right \rangle = \left \langle T^{\mu}_{\mu} \right\rangle$ 
and the generalized pressure is
\begin{equation}
    \Psi = {1 \over 3} \left(\left\langle T \right\rangle - 
    \Lambda\right)~.
\end{equation}
Similarly, one obtains the coefficients \cite{joy}
\begin{eqnarray}
    {\cal A} &=& \frac {- \left(n_{\mu} p^{\nu} \left\langle 
           T^{\mu}_{\nu} \right\rangle + x^2 \Lambda\right)} 
           {\left(x^4 - n^2 p^2\right)}~,\nonumber\\
       && \\
    {\cal B} &=& \frac {\left(p_{\mu} p^{\nu} \left\langle 
           T^{\mu}_{\nu} \right\rangle + p^2 \Lambda\right)}
           {\left(x^4 - n^2 p^2\right)}~,\nonumber\\
       && \\
    {\cal C} &=& \frac {\left(n_{\mu} n^{\nu} \left\langle 
           T^{\mu}_{\nu} \right\rangle + n^2 \Lambda\right)}
           {\left(x^4 - n^2 p^2\right)}~. \label{alg}
\end{eqnarray}

Relating neutron ($n^{\mu}$) and proton ($p^{\mu}$) number density currents to 
mean particle fluxes of neutrons and protons along the $z$ direction in the 
relativistic mean field (RMF) model \cite{joy}, 
it follows from Eq.({\ref{lcov}}) that
\begin{equation}
    \Lambda = \left \langle T^0_0 \right\rangle + \left \langle 
              T^z_z \right\rangle - \left\langle T^x_x \right\rangle~,
\end{equation}
where averaged stress-energy components are calculated in the RMF model and
those are defined below.

Next the implications of slow rotation are discussed in the following paragraph 
\cite{joy}. Here it is assumed that the space-time is flat in local regions of 
fluid elements. Since $x^2 - n p$ is small with respect to $n p$ \cite{joy}, 
this leads to the analytic expansion of the master function as
\begin{equation}
\Lambda(n^2,p^2,x^2) = \sum_{i = 0}^{\infty} \lambda_i(n^2,p^2) 
                           \left(x^2 - n p\right)^i~. 
\label{slow}
\end{equation}

Coefficients in the field equations are given by \cite{joy}, 
\begin{eqnarray}
  {\cal A} &=& - \sum_{i = 1}^{\infty} i~\lambda_i(n^2,p^2) \left(x^2 - n 
         p\right)^{i - 1}~, \cr
     && \cr
  {\cal B} &=& - {1 \over n} {\partial \lambda_0 \over \partial n} - {p 
         \over n} {\cal A} - {1 \over n} \sum_{i = 1}^{\infty} {\partial 
         \lambda_i \over \partial n} \left(x^2 - n p\right)^i~, \cr
     && \cr
  {\cal C} &=& - {1 \over p} {\partial \lambda_0 \over \partial p} - {n 
         \over p} {\cal A} - {1 \over p} \sum_{i = 1}^{\infty} {\partial 
         \lambda_i \over \partial p} \left(x^2 - n p\right)^i~. 
\end{eqnarray}

The master function is calculated within a RMF model 
\cite{joy,gle}. In this case, the relative motion between neutrons and protons
is taken into account. In the RMF model, nucleon-nucleon interaction is 
mediated by the exchange of mesons. Comer and Joynt \cite{joy,com04} made the 
connection 
between macroscopic fluid system and microscopic RMF model for the first time.
They used the relativistic $\sigma$-$\omega$ model in their calculation 
\cite{joy}. However, neutron star matter is highly isospin asymmetric matter. 
This can be taken care of by the inclusion of $\rho$ mesons in the RMF model.
We extend the calculation of Comer and Joynt \cite{joy} to include $\rho$ 
mesons as well as scalar meson self-interactions.
The Lagrangian density for nucleon-nucleon interaction 
is given by \cite{gle} 
\begin{eqnarray}\label{lag}
{\cal L}_B &=& \sum_{B=n,p} \bar\Psi_{B}\left(i\gamma_\mu{\partial^\mu} - m_B
+ g_{\sigma B} \sigma - g_{\omega B} \gamma_\mu \omega^\mu
- g_{\rho B}
\gamma_\mu{\mbox{\boldmath t}}_B \cdot
{\mbox{\boldmath $\rho$}}^\mu \right)\Psi_B\nonumber\\
&& + \frac{1}{2}\left( \partial_\mu \sigma\partial^\mu \sigma
- m_\sigma^2 \sigma^2\right) - U(\sigma) \nonumber\\
&& -\frac{1}{4} \omega_{\mu\nu}\omega^{\mu\nu}
+\frac{1}{2}m_\omega^2 \omega_\mu \omega^\mu
- \frac{1}{4}{\mbox {\boldmath $\rho$}}_{\mu\nu} \cdot
{\mbox {\boldmath $\rho$}}^{\mu\nu}
+ \frac{1}{2}m_\rho^2 {\mbox {\boldmath $\rho$}}_\mu \cdot
{\mbox {\boldmath $\rho$}}^\mu ~.
\end{eqnarray}
Here $\psi_B$ denotes the Dirac bispinor for baryons $B$ with vacuum mass $m_B$
and the isospin operator is ${\mbox {\boldmath t}}_B$. The scalar
self-interaction term \cite{bog} is
$U(\sigma)=\frac{1}{3}b m \left(g_\sigma \sigma\right)^3 
+ \frac{1}{4}c  \left(g_\sigma \sigma\right)^4$. The Dirac nucleon effective 
mass $m_*$ is defined as $m_*=m-<g_\sigma \sigma>$. 
Here we use the nucleon mass ($m$) which is the average of bare neutron 
($m_n$) and proton ($m_p$) masses. Further
we use $c_{\sigma}^{2}=(g_\sigma/m_\sigma)^2$ , 
$c_{\omega}^2=(g_\omega/m_\omega)^2$ and  $c_{\rho}^2=(g_\rho/m_\rho)^2$ in 
this calculation.

Here we adopt the mean field approximation to solve the equations of motion 
for meson fields \cite{gle}. We choose a frame in which neutrons have zero 
spatial momentum and protons have a wave vector $k_{\mu}=(k_0,0,0,K)$ 
\cite{joy}. We obtain the meson field equations as, 
    \begin{equation}\label{effm}
     m_*=m-c_{\sigma}^2 \left\langle {\bar{\psi}\psi} \right\rangle  + {b} {m} c_\sigma^2 \left(m-m_*\right)^2 + c c_\sigma^2\left(m-m_*\right)^3
    \end{equation} 
    \begin{equation}\label{omeg}
    {g_\omega \omega_0}= - c_\omega^{2}(n^0 + p^0)~,
     \end{equation}
    \begin{equation}\label{ome1}
    {g_\omega \omega^z}= - c_\omega^{2} (n^z + p^z)~,
    \end{equation}
    \begin{equation}\label{rho}
    {g_\rho {\rho_3^0}}= - \frac{1}{2}{c_\rho}^2({p^0}-{n^0})~,
    \end{equation}
     \begin{equation}
    { g_\rho {\rho^z_3}}= - \frac{1}{2}{c_\rho}^2({p^z}-{n^z})~,
    \end{equation}
where 
$p^0=\left\langle{\bar\psi_p \gamma^0\psi_p}\right\rangle=k_p^3/{3\pi^2}$,
$n^0=\left\langle{\bar\psi_n \gamma^0\psi_n}\right\rangle=k_n^3/{3\pi^2}$, 
$p^z=\left\langle{\bar\psi_p \gamma^z\psi_p}\right\rangle$, and 
$n^z=\left\langle{\bar\psi_n \gamma^z\psi_n}\right\rangle$.

In the zero momentum frame of neutrons, averaged stress-energy tensor components
are given by, 
\begin{eqnarray}
    \left\langle T^{0}_{0} \right\rangle &=& - {1 \over 2} 
    c_{\omega}^2 \sum_{B=n,p} \left(\left\langle \bar{\psi}_B \gamma^{0} \psi_B 
    \right\rangle^2 - \left\langle \bar{\psi}_B \gamma^{z} \psi_B 
    \right\rangle^2\right)  - {1 \over 2} 
c_{\rho}^2 \sum_{B=n,p}\left(\left\langle \bar{\psi}_B I_{3B} \gamma^{0} \psi_B 
    \right\rangle^2 - \left\langle \bar{\psi}_B I_{3B} \gamma^{z} \psi_B 
    \right\rangle^2\right)\nonumber \\
    && - {1 \over 2} c_{\sigma}^{- 2} \left(m^2 
    - m_*^2\right)-\frac{1}{3}b m \left(m-m_*\right)^3 -\frac{1}{4} c \left(m-m_*\right)^4
    - \sum_{B=n,p}\left\langle \bar{\psi}_B \gamma^i k_i \psi_B 
    \right\rangle \ ,\label{tensor1} \\
    && \cr
    \left\langle T^{0}_{z} \right\rangle &=& \sum_{B=n,p}\left\langle 
     \bar{\psi}_B 
    \gamma^{0} k_{z} \psi_B \right\rangle \ , \\
    && \cr
    \left\langle T^{x}_{x} \right\rangle &=& \left\langle T^{y}_{y} 
    \right\rangle = {1 \over 2} c_{\omega}^2 \sum_{B=n,p}\left(\left\langle 
    \bar{\psi}_B \gamma^{0} \psi_B \right\rangle^2 - \left\langle 
    \bar{\psi}_B \gamma^{z} \psi_B \right\rangle^2\right) + {1 \over 2} 
c_{\rho}^2 \sum_{B=n,p}\left(\left\langle \bar{\psi}_B I_{3B} \gamma^{0} \psi_B 
 \right\rangle^2 - \left\langle \bar{\psi}_B I_{3B}\gamma^{z} \psi_B 
    \right\rangle^2\right)\cr
    && \cr
    && - {1 \over 2} 
    c_{\sigma}^{- 2} \left(m - m_*\right)^2 -\frac{1}{3}b m \left(m-m_*\right)^3 -\frac{1}{4} c \left(m-m_*\right)^4
    + \sum_{B=n,p} \left\langle \bar{\psi}_B 
    \gamma^{x} k_{x} \psi_B \right\rangle \ , \\
    && \cr
    \left\langle T^{z}_{z} \right\rangle &=& {1 \over 2} 
    c_{\omega}^2 \sum_{B=n,p}\left(\left\langle \bar{\psi}_B \gamma^{0} \psi_B 
    \right\rangle^2 - \left\langle \bar{\psi}_B \gamma^{z} \psi_B 
    \right\rangle^2\right)+{1 \over 2} 
 c_{\rho}^2 \sum_{B=n,p}\left(\left\langle \bar{\psi}_B I_{3B}\gamma^{0} \psi_B 
 \right\rangle^2 - \left\langle \bar{\psi}_B I_{3B} \gamma^{z} \psi_B 
    \right\rangle^2\right)\nonumber \\
    &&- {1 \over 2} c_{\sigma}^{- 2} \left(m - 
    m_*\right)^2 -\frac{1}{3}b m \left(m-m_*\right)^3 -\frac{1}{4} c \left(m-m_*\right)^4
    + \sum_{B=n,p}\left\langle \bar{\psi}_B \gamma^{z} k_{z} \psi_B 
    \right\rangle \ ,\label{tensor2} 
\end{eqnarray}%
where $I_{3B}$ is the third isospin component for baryon $B$. Averaged 
stress-energy tensor components include terms which are to be integrated over 
neutron and proton Fermi surfaces.

We perform integrations in cylindrical coordinates with 
the definitions $\phi_{\omega} = g_{\omega} \omega^z$,
$\phi_{\omega K} = \phi_{\omega} + K$ 
and 
$\phi _\rho = g_{\rho} {\rho^z_3}$ as in Ref.{\cite{joy}}. We write the 
effective mass explicitly as,

\begin{eqnarray}
    m_* &=& m - {c_{\sigma}^2 \over 2 \pi^2}  m_* \left(
            \int_{- k_n}^{k_n} d k_z \left[k_n^2 + \phi_{\omega}^2 
            +{1\over 4}\phi_\rho^2 + m_*^2 + 2 \phi_{\omega} k_z
            -\phi_\rho k_z - \phi_{\omega} \phi_{\rho}\right]^{1/2} \right. \cr
         && \cr
         && \left.
            +\int_{- k_p}^{k_p} d k_z \left[k_p^2 + \left(
            \phi_{\omega K}+{1\over 2}\phi_\rho \right)^2 + m_*^2 + 2 \left(\phi_{\omega K}+{1\over 2}\phi_\rho \right) 
            k_z\right]^{1/2}  \right. \cr
         && \cr
         &&- \left.\int_{- k_n}^{k_n} d k_z \left[\left(k_z + \phi_{\omega}-{1\over 2}\phi_\rho
            \right)^2 + m_*^2\right]^{1/2} - \int_{- k_p}^{k_p} 
          d k_z \left[\left(k_z + \phi_{\omega K} +{1\over 2}\phi_\rho \right)^2 
            + m_*^2\right]^{1/2}\right) \cr 
         && \cr
         && + b m c_\sigma^2 \left(m-m_*\right)^2 
            + c c_\sigma^2 \left(m-m_*\right)^3~. 
\label{efm}
 \end{eqnarray}

The z components of neutron and proton number current densities take the form         
\begin{eqnarray}         
    n^{z} &=& {1 \over 2 \pi^2} \int_{- k_{n}}^{k_{n}} d k_{z} 
               \left(k_{z} + \phi_{\omega}-{1\over 2}\phi_\rho\right) \left(\left[
               k_{n}^2 + m_*^2 + \phi_{\omega}^2-{1\over 4}\phi_\rho^2 
               + 2 \phi_{\omega} k_{z}-\phi_\rho k_z
               \right]^{1/2}  \right. \cr
         && \cr
         && \left.
         - \left[\left(k_{z} + \phi_{\omega}-{1\over 2}\phi_\rho\right)^2 
               + m_*^2\right]^{1/2}\right)~, \cr
         && \cr
    p^{z} &=& {1 \over 2 \pi^2} \int_{- k_{p}}^{k_{p}} d k_{z} 
               \left(k_{z} + \phi_{\omega K}+{1\over 2}\phi_\rho \right) 
                \left(\left[k_{p}^2 + m_*^2 + \left(\phi_{\omega K} 
              +{1\over 2}\phi_\rho \right)^2 + 
               2 \left(\phi_{\omega K}+{1\over 2}\phi_\rho 
              \right) k_{z}\right]^{1/2} \right. \cr
            && \cr
            && - \left.\left[\left(k_{z} + \phi_{\omega K}
              +{1\over 2}\phi_\rho \right)^2 
               + m_*^2\right]^{1/2}\right)~. 
\label{app}
\end{eqnarray}

The master function in Eq.({\ref{lcov}}) can be written as \cite{com04}
\begin{eqnarray}
\Lambda &= & - {c_{\omega}^2 \over {18 \pi^4}}\left(k^3_n + k^3_p\right)^2
             -{c_{\rho}^2 \over {72 \pi^4}}\left(k^3_p - k^3_n\right)^2- 
        \frac{1}{2 c_{\omega}^2}\phi_{\omega}^2- \frac{1}{2 c_{\rho}^2}\phi_\rho^2\nonumber \\
             && - \frac{1}{2 c_{\sigma}^2}(m^2-m_*^2)-\frac{1}{3} b m 
           \left(m - m_*\right)^3 -\frac{1}{4} c \left(m-m_*\right)^4\nonumber\\
             && - 3 \sum_{B=n,p}\left\langle \bar{\psi_B}\gamma^x k_x\psi_B\right\rangle~,
             \end{eqnarray}
where,
\begin{eqnarray}{\hspace*{-1.5cm}}
\sum_{B=n,p}\left\langle \bar{\psi_B} \gamma^{x} k_{x} \psi_B \right\rangle &=& 
    {1 \over 12 \pi^2} \left(
    \int_{- k_n}^{k_n} d k_z \left[(k_n^2 - 2 m_*^2 - 2 \phi_{\omega}^2 
    - {1\over 2}\phi_{\rho}^2 -3 k_z^2 - 4 \phi_{\omega} k_z + 2 \phi_{\rho} k_z
      + 2 \phi_{\omega} \phi_{\rho}) \right.\right. \cr 
         && \cr
         && \left.\left.
(k_n^2 + \phi_{\omega}^2 +{1 \over 4} \phi_{\rho}^2 + m_*^2 + 2 \phi_{\omega}k_z
   -\phi_{\rho}k_z - \phi_{\omega} \phi_{\rho})^{1/2} + 2([k_z + \phi_{\omega} 
   - {1 \over 2} \phi_{\rho}]^2 + m_*^2)^{3/2}\right] \right. \cr
         && \cr
         && \left.
      +  \int_{- k_p}^{k_p} d k_z \left[(k_p^2 - 2 m_*^2 - 2 \phi_{\omega K}^2 
  - {1\over 2}\phi_{\rho}^2 -3 k_z^2 - 4 \phi_{\omega K} k_z - 2 \phi_{\rho} k_z
      - 2 \phi_{\omega K} \phi_{\rho}) \right.\right. \cr 
         && \cr
         && \left.\left.
      (k_p^2 + \phi_{\omega K}^2 +{1 \over 4} \phi_{\rho}^2 + m_*^2
     + 2\phi_{\omega K}k_z +\phi_{\rho}k_z + \phi_{\omega K} \phi_{\rho})^{1/2}
     + 2([k_z + \phi_{\omega K} 
   + {1 \over 2} \phi_{\rho}]^2 + m_*^2)^{3/2}\right]\right).\nonumber\\
             \end{eqnarray}

In evaluating the master function as well as the coefficients, the slow rotation
approximation which implies that $K$ should be small compared with $k_{n,p}$, is
exploited as detailed in the Appendix. We are dealing with superfluidity in
neutron star matter which is made of neutrons, protons and electrons. When we
neglect the relative motion between neutron and proton fluids, $-\Lambda|_0$ 
becomes the energy density of the neutron star matter. We add the contribution 
of electrons to the master function ($\Lambda$). Here, electrons are treated as 
noninteracting relativistic particles. Therefore, in the
limit $K \rightarrow 0$, the master function
which is the first term of Eq.(\ref{slow}), 
generalized pressure and the chemical potentials of neutron and 
proton fluids are given by 
\begin{eqnarray}
    \left.\Lambda\right|_0 &=&  - {c_\omega^2\over {18 \pi^4}}\left(k_n^3+k_p^3\right)^2-{c_\rho^2\over {72 \pi^4}}\left(k_p^3-k_n^3\right)^2- {1 \over 4 \pi^2} \left(k_n^3 
             \sqrt{k_n^2 + \left.m^2_*\right|_0} + 
             k_p^3 \sqrt{k_p^2 + \left.m^2_*\right|_0}
             \right)   \cr
             && \cr
             &&- {1 \over 
             4} c_\sigma^{- 2} \left[\left(2 m - \left.m_*\right|_0\right) 
             \left(m - \left.m_*\right|_0\right)+\left.m_*\right|_0\left( b m c_\sigma^2\left(m-\left.m_*\right|_0\right)^2 
          + c  c_\sigma^2\left(m-\left.m_*\right|_0\right)^3\right)\right]\nonumber \\
          &&-{1 \over 8 \pi^2} \left( k_p \left[2 k_p^2 + m_e^2
             \right] \sqrt{k_p^2 + m^2_e} - m^4_e {\rm ln}\left[
             {k_p + \sqrt{k_p^2 + m^2_e} \over m_e}\right]\right) 
             \ , \\
             && \cr
    \left.\mu\right|_0 &=&-\frac{\pi^2}{k_n^2} \left.\frac{\partial \Lambda}{\partial k_n}\right|_0={c_\omega^2\over {3 \pi^2}}\left(k_n^3+k_p^3\right)
 - {c_\rho^2\over {12 \pi^2}}\left(k_p^3-k_n^3\right)  + \sqrt{k_n^2 + 
             \left.m^2_*\right|_0} \ , \\ 
             && \cr
    \left.\chi\right|_0 &=&-\frac{\pi^2}{k_p^2} \left.\frac{\partial \Lambda}{\partial k_p}\right|_0= {c_\omega^2\over {3 \pi^2}}\left(k_n^3+k_p^3\right)
+ {c_\rho^2\over {12 \pi^2}}\left(k_p^3-k_n^3\right) + \sqrt{k_p^2 + 
             \left.m^2_*\right|_0} + \sqrt{k_p^2 + m_e^2} \  , \\ 
             && \cr
    \left.\Psi\right|_0 &=& \left.\Lambda\right|_0 + {1 \over 3 \pi^2}
             \left(\left.\mu\right|_0 k_{n}^3 + 
             \left.\chi\right|_0 k_{p}^3\right)~,
\end{eqnarray}
where the subscript "0" stands for quantities calculated in the limit 
$K \rightarrow 0$. It is to be noted here that energy density $-\Lambda|_0$ 
and pressure $\Psi|_0$ constitute the equation of state for the calculation of
equilibrium configurations of neutron stars which we discuss in Sec. III.

Coefficients in momentum covectors are given by, 
\begin{eqnarray}
{\cal A}|_0 &=& c_{\omega}^2-\frac{1}{4} c_{\rho}^2 + {c^2_{\omega} \over 5 
        \left.\mu^2\right|_0} \left(2 k_p^2 {\sqrt{k_n^2 + 
        \left.m^2_*\right|_0} \over \sqrt{k_p^2 + 
        \left.m^2_*\right|_0}} + {c^2_{\omega} \over 3 \pi^2} 
        \left[{k_n^2 k_p^3 \over \sqrt{k_n^2 + 
        \left.m^2_*\right|_0}} + {k_p^2 k_n^3 \over \sqrt{k_p^2 + 
        \left.m^2_*\right|_0} }\right]\right)\cr
         && \cr
         &&
        +{c^2_{\rho} \over 20 
        \left.\mu^2\right|_0} \left(2 k_p^2 {\sqrt{k_n^2 + 
        \left.m^2_*\right|_0} \over \sqrt{k_p^2 + 
        \left.m^2_*\right|_0}} + {c^2_{\rho} \over 12 \pi^2} 
        \left[{k_n^2 k_p^3 \over \sqrt{k_n^2 + 
        \left.m^2_*\right|_0}} + {k_p^2 k_n^3 \over \sqrt{k_p^2 + 
        \left.m^2_*\right|_0} }\right]\right)\cr
         && \cr 
         &&
         -{c^2_{\rho}c^2_{\omega} \over 30\left.\mu^2\right|_0 \pi^2} 
        \left[{k_n^2 k_p^3 \over \sqrt{k_n^2 + 
        \left.m^2_*\right|_0}} - {k_p^2 k_n^3 \over \sqrt{k_p^2 + 
        \left.m^2_*\right|_0} }\right]  + {3 \pi^2 k_p^2 
        \over 5 \left.\mu^2\right|_0 k_n^3} {k_n^2 + 
        \left.m^2_*\right|_0 \over \sqrt{k_p^2 + 
        \left.m^2_*\right|_0}} \ , \\
        && \cr
{\cal B}|_0 &=& {3 \pi^2 \left.\mu\right|_0 \over k_n^3} - 
        c_{\omega}^2 {k_p^3 \over k_n^3}+ \frac{1}{4}c_{\rho}^2 {k_p^3 \over k_n^3} - {c^2_{\omega} k_p^3 
        \over 5 \left.\mu^2\right|_0 k_n^3} \left(2 k_p^2 
        {\sqrt{k_n^2 + \left.m^2_*\right|_0} \over \sqrt{k_p^2 + 
        \left.m^2_*\right|_0}} + {c^2_{\omega} \over 3 \pi^2} 
        \left[{k_n^2 k_p^3 \over \sqrt{k_n^2 + 
        \left.m^2_*\right|_0}} + {k_p^2 k_n^3 \over \sqrt{k_p^2 + 
        \left.m^2_*\right|_0} }\right]\right)  \cr
        && \cr
        && - {c^2_{\rho} k_p^3 
        \over 20 \left.\mu^2\right|_0 k_n^3} \left(2 k_p^2 
        {\sqrt{k_n^2 + \left.m^2_*\right|_0} \over \sqrt{k_p^2 + 
        \left.m^2_*\right|_0}} + {c^2_{\rho} \over 12 \pi^2} 
        \left[{k_n^2 k_p^3 \over \sqrt{k_n^2 + 
        \left.m^2_*\right|_0}} + {k_p^2 k_n^3 \over \sqrt{k_p^2 + 
        \left.m^2_*\right|_0} }\right]\right)  \cr
        && \cr
        && +{c^2_{\rho} c^2_{\omega}k^3_p\over 30 \pi^2\left.\mu^2\right|_0 k_n^3} 
        \left[{k_n^2 k_p^3 \over \sqrt{k_n^2 + 
        \left.m^2_*\right|_0}} - {k_p^2 k_n^3 \over \sqrt{k_p^2 + 
        \left.m^2_*\right|_0} }\right]-{3 \pi^2 k_p^5 \over 5 \left.\mu^2\right|_0 k_n^6} 
        {k_n^2 + \left.m^2_*\right|_0 \over \sqrt{k_p^2 + 
        \left.m^2_*\right|_0}}\,\label{b00} \ , \\
        && \cr
{\cal C}|_0 &=& {3 \pi^2 \left.\chi\right|_0 \over k_p^3}+ \frac{1}{4}c_{\rho}^2 {k_n^3 \over k_p^3} - 
        c_{\omega}^2 {k_n^3 \over k_p^3} - {c^2_{\omega} k_n^3 
        \over 5 \left.\mu^2\right|_0 k_p^3} \left(2 k_p^2 
        {\sqrt{k_n^2 + \left.m^2_*\right|_0} \over \sqrt{k_p^2 + 
        \left.m^2_*\right|_0}} + {c^2_{\omega} \over 3 \pi^2} 
        \left[{k_n^2 k_p^3 \over \sqrt{k_n^2 + 
        \left.m^2_*\right|_0}} + {k_p^2 k_n^3 \over \sqrt{k_p^2 + 
        \left.m^2_*\right|_0} }\right]\right)  \cr
        && \cr
        && - {c^2_{\rho} k_n^3 
        \over 20 \left.\mu^2\right|_0 k_p^3} \left(2 k_p^2 
        {\sqrt{k_n^2 + \left.m^2_*\right|_0} \over \sqrt{k_p^2 + 
        \left.m^2_*\right|_0}} + {c^2_{\rho} \over 12 \pi^2} 
        \left[{k_n^2 k_p^3 \over \sqrt{k_n^2 + 
        \left.m^2_*\right|_0}} + {k_p^2 k_n^3 \over \sqrt{k_p^2 + 
        \left.m^2_*\right|_0} }\right]\right)\cr
        && \cr
        && +{c^2_{\rho} c^2_{\omega}k_n^3\over 30 \pi^2\left.\mu^2\right|_0 k_p^3} 
        \left[{k_n^2 k_p^3 \over \sqrt{k_n^2 + 
        \left.m^2_*\right|_0}} - {k_p^2 k_n^3 \over \sqrt{k_p^2 + 
        \left.m^2_*\right|_0} }\right]-{3 \pi^2 \over 5 \left.\mu^2\right|_0 k_p} 
        {k_n^2 + \left.m^2_*\right|_0 \over \sqrt{k_p^2 + 
        \left.m^2_*\right|_0}}~.
\label{coo} 
 \end{eqnarray}
Similarly, other coefficients which enter into the calculation of equilibrium
neutron star configurations, are calculated according to Ref.\cite{joy} and
given by, 

\begin{eqnarray}
{{\cal A}_0^0}|_0 &=-& \frac{\pi^4}{k_p^2k_n^2} \left.\frac{\partial^2 \Lambda}{\partial k_p\partial k_n}\right|_0
        = c_\omega^2 - \frac{c_\rho^2}{4}+ {\pi^2 \over k^2_p} { 
        \left.m_*\right|_0 \left.{\partial m_* \over \partial k_p}
        \right|_0 \over \sqrt{k^2_n + \left.m^2_*\right|_0}}\ , \\
        && \cr
{{\cal B}_0^0}|_0 &=& \frac{\pi^4}{k_n^5} \left(\left.2\frac{\partial \Lambda}{\partial k_n}\right|_0-k_n\left.\frac{\partial^2 \Lambda}{\partial k_n^2}\right|_0\right) 
        = c_\omega^2 + \frac{c_\rho^2}{4} + {\pi^2 \over k^2_n} {k_n + 
        \left.m_*\right|_0 \left.{\partial m_* \over \partial k_n}
        \right|_0 \over \sqrt{k^2_n + \left.m^2_*\right|_0}} \ , \\
        && \cr
{{\cal C}_0^0}|_0 &=& \frac{\pi^4}{k_p^5} \left(\left.2\frac{\partial \Lambda}{\partial k_p}\right|_0-k_p\left.\frac{\partial^2 \Lambda}{\partial k_p^2}\right|_0\right)
        = c_\omega^2 +  \frac{c_\rho^2}{4} + {\pi^2 \over k^2_p} {k_p + 
        \left.m_*\right|_0 \left.{\partial m_* \over \partial k_p}
        \right|_0 \over \sqrt{k^2_p + \left.m^2_*\right|_0}} + 
        {\pi^2 \over k_p} {1 \over \sqrt{k^2_p + m^2_e}}~.
\end{eqnarray}
Derivatives of the effective mass with respect to neutron and proton
Fermi momenta are explicitly shown in the Appendix. 

We obtain entrainment matrix elements inverting Eqs. (3) and (4) and compare
those with the relativistic analog of the mass density matrix ($\rho_{ik}$) 
\cite{joy,bash},
\begin{eqnarray} 
Y_{nn}&=&\frac{\rho_{nn}} {m^2}=\frac{{\cal C}|_0}{({\cal B}|_0 {\cal C}|_0
-{\cal A}|_0^2)}~,\nonumber\\ 
Y_{np}&=&\frac{\rho_{np}} {m^2}=-\frac{{\cal A}|_0}{({\cal B}|_0 {\cal C}|_0
-{\cal A}|_0^2)}~,\nonumber\\ 
Y_{pp}&=&\frac{\rho_{pp}} {m^2}=\frac{{\cal B}|_0}{({\cal B}|_0 {\cal C}|_0
-{\cal A}|_0^2)}~.
\label{element}
 \end{eqnarray}
The entrainment matrix in this form can be compared with that of 
Ref.\cite{gus09}. It is worth noting here that when Eqs. (3) and (4) are
inverted and neutron and proton number density currents are written in terms of 
chemical potentials, we obtain the relation ${\sum_{k=n,p}} Y_{ik} \mu_k = n_i$,
where $i=n,p$. 

Now the entrainment parameter ($\epsilon_{mom}$) in the zero momentum frame of 
neutrons is related to the off-diagonal component of the mass density matrix 
i.e.
$\rho_{np} = - \epsilon_{mom} m n$ and can be calculated 
in terms of coefficients in momentum covectors in two fluid formalism 
from the following relation \cite{joy},
\begin{equation}
\epsilon_{mom}=\frac{m}{n}\frac{{\cal A}|_0}{({\cal B}|_0 {\cal C}|_0
-{\cal A}|_0^2)}~. 
\end{equation}
Similarly, the entrainment parameter in the zero velocity frame of neutrons is
given by \cite{prix,joy},
\begin{equation}
\epsilon_{vel} = \frac{{{\cal A}|_0}n}{m}~.
\end{equation}
In the nonrelativistic case, an explicit relationship between the entrainment
parameter and effective nucleon mass was found by various groups 
\cite{prix,haen}.  

Finally, charge neutrality and $\beta$-equilibrium conditions are to be imposed
in neutron star matter. The charge neutrality condition is $k_p = k_e$. The 
condition of chemical equilibrium for $npe$ matter 
is $\left.\mu\right|_0 = \left.\chi\right|_0$, where
$\mu|_0$ and $\chi|_0$ are the neutron and proton plus electron chemical 
potentials, respectively. 

\section{Results and Discussion}
Meson-nucleon couplings $c_{\sigma}$, $c_{\omega}$, $c_{\rho}$, $b$ and $c$ of
the Lagrangian density in Eq. (14) are 
determined by reproducing nuclear matter saturation properties such as
binding energy per nucleon (-16.3 MeV), saturation density 
($n_0 = 0.153$ fm$^{-3}$), Dirac 
nucleon effective mass ($m_*/m = 0.7$), the symmetry energy coefficient 
(32.5 MeV)
and incompressibility (200 MeV). These coupling constants are taken from
the Ref.\cite{gle}. This parameter set is known as the GL 
set. We also perform the calculation using the non-linear (NL3) interaction 
\cite{ring}. New 
parametrization of the NL3 interaction reproducing binding energy per nucleon 
(-16.24 MeV), saturation density (0.148 fm$^{-3}$), incompressibility 
(271.5 MeV), the symmetry energy coefficient (37.29 MeV) and the slope of the
symmetry energy (118.2 MeV) \cite{horo}, is adopted in our 
calculation. Both parameter sets are listed in Table I.  

In this calculation, we consider the $\beta$-equilibrated neutron star matter 
made of neutrons ($n$), protons ($p$) and electrons ($e$). 
Equilibrium configurations of neutron stars are calculated 
following the prescription of Comer and other collaborators \cite{joy,lan99} 
and using our equation of state as given by energy density ($-\Lambda|_0$) and
pressure ($\Psi|_0$) in Eqs. (28) and (31), respectively.  
Neutron and proton Fermi momenta at the center of the star are needed for this 
purpose. For a given value of neutron 
Fermi momentum or wave number, proton Fermi momentum is calculated from the
$\beta$-equilibrium condition. We perform this calculation for the GL and NL3
parameter sets. Neutron star masses as a function of central 
neutron density for NL3 (solid line) and GL (dashed line) sets are plotted 
in Fig. 1. It is noted that maximum neutron star masses corresponding to the 
GL and NL3 parameters are well above the observed limit of 2.01$\pm 0.04$ 
$M_{\odot}$ \cite{anto}. For the calculation of entrainment, we choose neutron 
star configurations which are just below maximum masses in both cases.
In the case of the GL set, we consider a neutron star mass of 2.37 
$M_{\odot}$ corresponding to the central value of neutron wave number 
$k_n (0) = 2.71 fm^{-1}$ and proton fraction 0.24. The radius of
the neutron star is 11.09 km. Similarly, in the other
case with the NL3 set, we find a neutron star having maximum mass 2.82 
$M_{\odot}$ and radius 13.17 km.
The corresponding central values of neutron wave number and proton fraction are
2.40 fm$^{-1}$ and 0.23, respectively. 

We calculate dynamical neutron and proton effective masses \cite{cha08} using 
$\bar m_*^n = n_n{\cal B}|_0$ and  $\bar m_*^p = n_p{\cal C}|_0$, where $n_n$
and $n_p$ are neutron and proton number densities. Dynamical effective masses 
are
plotted as a function of baryon density in Fig. 2. The left panel shows the
results of the GL parameter set whereas the right panel denotes those of the
NL3 parameter set. In both panels, the upper curve represents
the neutron effective mass and the lower curve corresponds to the proton 
effective mass. It is noted that the neutron effective mass increases with
density and becomes greater than the free neutron mass in both cases. However, 
it rises faster in the NL3 case. On the other hand, the 
proton effective mass decreases with density initially and rises at higher 
densities. However, its value always stays below the free proton mass. In the GL
case, the proton effective mass is always higher than that of the NL3 case. 
These findings are different from the effective masses calculated in the
nonrelativistic calculations as noted already in Refs.\cite{cha08,haen}.    

We also calculate the Landau effective mass for nucleons and it is related to 
the Dirac effective mass through the expression 
$m^{*i}_L = \sqrt{{k_i}^2 + (m - g_{\sigma} \sigma)^2}$ \cite{gus09}. 
Landau effective masses
for neutrons and protons are shown as a function of baryon density in Fig. 3.
The left panel shows the results of the GL set and the right panel represents 
those of the NL3 set. 
Here we find that neutron and proton effective masses decrease as baryon 
density increases in both panels and they are below their bare masses. It is 
noted
that the Landau effective masses are always higher in the GL set than those of
the NL3 set.

Normalsied entrainment matrix elements of Eq.({\ref{element}}) are 
shown as a function of baryon density in Fig. 4. The normalisation constant 
is chosen as $Y = 3n_0/{\mu_{n}(3 n_0)}$ as it was done in Ref.\cite{gus09}. 
Entrainment matrix elements obtained in this calculation are compared with 
those calculated in the relativistic Landau Fermi liquid theory
\cite{gus09}. Both calculations are performed using the GL set of Table I.
Solid lines represent the results of Ref. \cite{gus09} whereas dashed lines
demonstrate the results of Eq. (\ref{element}) using the GL set. 
Though the results
of the two calculations are qualitatively similar, those are quantitatively 
very different. The difference between the results of the two calculations is 
negligible initially because the entrainment effect becomes small 
in the low density region 
in both approaches. On the other hand, this difference grows at 
higher baryon densities as the entrainment effect becomes more dominant in our 
case than that of 
Ref. \cite{gus09}. This may be attributed to  different formalisms in two 
calculations .
We also compare 
normalised matrix elements calculated in our model using the GL and NL3 
parameter sets in 
Fig. 5. Results of the NL3 set are higher than those of the GL set. The 
difference in the equations of state for the two parameter sets 
is reflected in the results of matrix element calculations. 

Next we present the results of entrainment parameters in the zero momentum and
zero velocity frames of neutrons. The radial profiles of the 
entrainment 
parameter in the zero momentum frame for the GL and NL3 sets 
are shown in Fig. 6.  These radial profiles are obtained for neutron stars of
mass $M = 2.37 M_{\odot}$ and radius $R= 11.09$ km in the case of the GL set
and $M = 2.82 M_{\odot}$ and radius $R= 13.17$ km in the case of the NL3 set.
The entrainment parameter in both cases remains constant in the core and drops 
rapidly at the surface. We find an appreciable difference between the two 
results 
towards the center. Moreover, in both cases, the entrainment effect is strong at
higher
baryon densities in the core whereas this effect diminishes sharply at lower
densities towards the surface. The value of the
entrainment parameter lies in the physical range $0 \le \epsilon_{mom} \le 1$
as found in earlier calculations \cite{prix,haen}. We compare this result with
that of the situation excluding $\rho$ mesons which was actually studied in 
Ref. \cite{joy}, as displayed in Fig. 7 for the
GL set. For the
calculation of the entrainment parameter without $\rho$ mesons, we obtain the 
radial profile of the entrainment parameter in a neutron star of mass 
2.33 $M_{\odot}$ and radius 10.96 km. It is evident from Figs. 6 and Fig. 7 that
the inclusion of $\rho$ in the calculation strongly enhances the entrainment 
parameter.

Further the radial profile of the entrainment parameter ($\epsilon_{vel}$) in 
the zero velocity frame is exhibited in Fig. 8. The solid line denotes the 
calculation without $\rho$ mesons and the dashed line implies the case including
$\rho$
mesons. It is noted that the entrainment parameter calculated without $\rho$ 
mesons is larger compared with the entrainment parameter with $\rho$ meson. 
This 
finding is opposite to what we see in the calculation of the entrainment 
parameter in the zero momentum frame. We also find that the values of 
the entrainment parameter
in the zero velocity frame are higher than those of the entrainment parameter in
the zero momentum frame.
Finally, we compare the radial profiles of entrainment parameters in the 
zero velocity frame for the GL and NL3 parameter sets as shown in Fig. 9. It
is evident from the figure that the two results do not differ much. 

So far we have neglected muons in our calculation. However, muons can be 
populated in neutron star matter when the threshold condition involving electron
and muon chemical potentials, $\mu_e = \mu_{\mu}$ is satisfied. We repeat our 
calculation including muons. However, muons have negligible effects on the
entrainment matrix elements and entrainment parameter.

\section{Summary and Conclusions}
We have extended the calculation of Comer and Joynt \cite{joy} to include $\rho$
mesons and the self-interaction term in the RMF model. Here we calculate
entrainment matrix elements and entrainment parameters using this model and 
the GL and NL3 parameter sets.
It is noted that the entrainment parameter in the zero momentum frame is 
significantly enhanced due to the presence of $\rho$ mesons in the calculation.
Furthermore we compare our results with those of the relativistic Landau Fermi 
liquid theory \cite{gus09} and find appreciable differences. 

Our calculation may be extended to include hyperons in a straightforward manner 
using a three fluid description \cite{andc} and applied to study the dynamics 
of superfluid
neutron stars. This could be compared with the findings of earlier calculations
including hyperons in the relativistic Landau Fermi liquid theory 
\cite{gus09,gus092}.

\section{Appendix}
In the slow rotation approximation, we expand scalar and vector quantities in 
terms of $K$. Scalar quantities like $m_*$ and $\Lambda$ depend on even powers
of $K$ whereas vector quantities depend on odd powers
of $K$. We keep terms up to $K^2$ in our calculation. Effective mass, 
$z$ components of $\omega$ and $\rho$ fields are expanded in the 
following way \cite{joy},
\begin{eqnarray}
\phi_{\omega}&=&\left.{\partial \phi_{\omega} \over \partial K}\right|_0 K~,\nonumber\\ 
\phi_{\rho}&=&\left.{\partial \phi_{\rho}\over \partial K}\right|_0 K~,\nonumber\\ 
m_*&=&\left.m_*\right|_0 + \left.{\partial m_* \over \partial K^2}\right|_0 K^2~.\nonumber\\ 
\label{expa}
\end{eqnarray}
Here,
\begin{eqnarray}
    \left.m_*\right|_0 &=& m_*(k_n,k_p,0) \cr
         && \cr
        &=& m - \left.m_*\right|_0 {c_\sigma^2 \over 2 \pi^2}  
            \left(k_n \sqrt{k_n^2 + \left.m^2_*\right|_0} + k_p 
            \sqrt{k_p^2 + \left.m^2_*\right|_0} + {1 \over 2} 
            \left.m_*^2\right|_0 {\rm ln} \left[{- k_n + 
            \sqrt{k_n^2 + \left.m^2_*\right|_0} \over k_n + 
            \sqrt{k_n^2 + \left.m^2_*\right|_0}}\right] \right. \cr
         && \cr
         && + {1 \over 2} \left.\left.m_*^2\right|_0 {\rm ln} \left[{- 
            k_p + \sqrt{k_p^2 + \left.m^2_*\right|_0} \over k_p + 
            \sqrt{k_p^2 + \left.m^2_*\right|_0}}\right]\right) \ + b  m  c_\sigma^2\left(m - m_*\right)^2 +   c  c_\sigma^2\left(m - m_*\right)^3~. 
            \label{mstar}
\end{eqnarray}

Plugging Eqs.({\ref{expa}}) in Eq.(\ref{efm}) and Eq.(\ref{app}) and expanding 
and keeping terms up to orders $K^2$, we obtain
\begin{eqnarray}
     \left.{\partial m_* \over \partial k_n}\right|_0 &=& - 
          {c_\sigma^2 \over \pi^2} {\left.m_*\right|_0 k_n^2 
          \over \sqrt{k_n^2 + \left.m^2_*\right|_0}} \left({3 m - 2 
          \left.m_*\right|_0 +3 b m c_\sigma^2\left(m-\left.m_*\right|_0\right)^2 
          +3 c  c_\sigma^2\left(m-\left.m_*\right|_0\right)^3\over\left.m_*\right|_0}\right.\nonumber \\ 
          &&\left.- {c_\sigma^2 
          \over \pi^2} \left[{k_n^3 \over \sqrt{k_n^2 + \left.m^2_*
          \right|_0}} + {k_p^3 \over \sqrt{k_p^2 + 
          \left.m^2_*\right|_0}}\right]+2 b m c_\sigma^2\left(m-\left.m_*\right|_0\right) 
          +3 c  c_\sigma^2\left(m-\left.m_*\right|_0\right)^2\right)^{- 1}~, 
          \\
          && \cr
     \left.{\partial m_* \over \partial k_p}\right|_0 &=& - 
          {c_\sigma^2 \over \pi^2} {\left.m_*\right|_0 k_p^2 
          \over \sqrt{k_p^2 + \left.m^2_*\right|_0}} \left({3 m - 2 
          \left.m_*\right|_0 +3 b m c_\sigma^2\left(m-\left.m_*\right|_0\right)^2 
          +3 c  c_\sigma^2\left(m-\left.m_*\right|_0\right)^3\over\left.m_*\right|_0}\right.\nonumber \\ 
          &&\left.- {c_\sigma^2 
          \over \pi^2} \left[{k_n^3 \over \sqrt{k_n^2 + \left.m^2_*
          \right|_0}} + {k_p^3 \over \sqrt{k_p^2 + 
          \left.m^2_*\right|_0}}\right]+2 b m c_\sigma^2\left(m-\left.m_*\right|_0\right) 
          +3 c  c_\sigma^2\left(m-\left.m_*\right|_0\right)^2\right)^{- 1} 
          ~, 
          \\
          && \cr    
      n^{z} &=& {1 \over 3 \pi^2} {k_n^3 \over \sqrt{k_n^2 + 
          \left.m^2_*\right|_0}} \left(\left.{\partial \phi_{\omega} \over 
          \partial K}\right|_0 K-\frac{1}{2} \left.{\partial \phi_\rho \over 
          \partial K}\right|_0 K \right) , \\
          && \cr
      p^{z} &=& {1 \over 3 \pi^2} {k_p^3 \over \sqrt{k_p^2 + 
             \left.m^2_*\right|_0}} \left(\left.{\partial \phi_{\omega} 
             \over \partial K}\right|_0 K +\frac{1}{2} \left.{\partial \phi_\rho \over 
          \partial K}\right|_0 K + K\right)  \ . 
\end{eqnarray}
and also
\begin{equation}
p^z + n^z=-\frac{1}{c_\omega^2}\left.{\partial \phi_{\omega} \over \partial K }\right|_0 K
\end{equation}
\begin{equation}
p^z - n^z=-\frac{2}{c_\rho^2}\left.{\partial \phi_\rho \over \partial K }\right|_0 K
\end{equation}
Using the four equations above, we get
\begin{eqnarray}
  \left.{\partial \phi_z \over \partial K}\right|_0 &=& - 
          {c_\omega^2 \over 3 \pi^2} {k_p^3 \over \sqrt{k_p^2 + 
          \left.m^2_*\right|_0}}\left(1+\frac{1}{4} {c_\rho^2 \over 3 \pi^2} {2 k_n^3 \over \sqrt{k_n^2 + 
          \left.m^2_*\right|_0}}\right)\over {\left(1 + {c_\omega^2+\frac{c_\rho^2}{4} \over 3 \pi^2} 
          \left[{k_n^3 \over \sqrt{k_n^2 + \left.m^2_*\right|_0}} + 
          {k_p^3 \over \sqrt{k_p^2 + \left.m^2_*\right|_0}}\right]
          +{c_\omega^2{c_\rho^2} \over 9 \pi^4} 
          \left[{k_n^3 k_p^3 \over \sqrt{(k_n^2 + \left.m^2_*\right|_0 ) 
           (k_p^2 + \left.m^2_*\right|_0)}}\right]\right)} \ , \\
          && \cr
        \frac{1}{2}\left.{\partial \phi_\rho \over \partial K}\right|_0 &=& - \frac{1}{4}
          {c_\rho^2 \over 3 \pi^2} {k_p^3 \over \sqrt{k_p^2 + 
          \left.m^2_*\right|_0}}\left(1+ {c_\omega^2 \over 3 \pi^2} {2 k_n^3 \over \sqrt{k_n^2 + 
          \left.m^2_*\right|_0}}\right)\over {\left(1 + {c_\omega^2+\frac{c_\rho^2}{4} \over 3 \pi^2} 
          \left[{k_n^3 \over \sqrt{k_n^2 + \left.m^2_*\right|_0}} + 
          {k_p^3 \over \sqrt{k_p^2 + \left.m^2_*\right|_0}}\right]
          +{c_\omega^2{c_\rho^2} \over 9 \pi^4} 
          \left[{k_n^3 k_p^3\over \sqrt{(k_n^2 + \left.m^2_*\right|_0)  
           (k_p^2 + \left.m^2_*\right|_0)}}\right]\right)} \ .
          \end{eqnarray}

Further we find
$\left.{\partial m_* \over \partial K^2}\right|_0 = 0~.$

\newpage
\begin{table}
\caption{Nucleon-meson coupling constants in 
the GL and NL3 sets are taken from Refs.\cite{gle,horo}. 
The coupling constants are obtained by reproducing the saturation properties
of symmetric nuclear matter as detailed in the text. 
All the parameters are in fm$^{2}$, except $b$ and $c$ which are dimensionless.}

\begin{center}
\vspace{5cm}
\begin{tabular}{cccccc} 

\hline\hline
\hfil& 
$c_{\sigma}^2$& $c_{\omega}^2$& $c_{\rho}^2$& $b$& $c$ \\ \hline
GL& 12.684&  7.148& 4.410& 0.005610& -0.006986 \\ \hline
NL3& 15.739& 10.530& 5.324& 0.002055& -0.002650 \\ \hline
\hline

\end{tabular}
\end{center}
\end{table}

\vspace{5cm}

\newpage 

\vspace{-5cm}

{\centerline{
\epsfxsize=12cm
\epsfysize=14cm
\epsffile{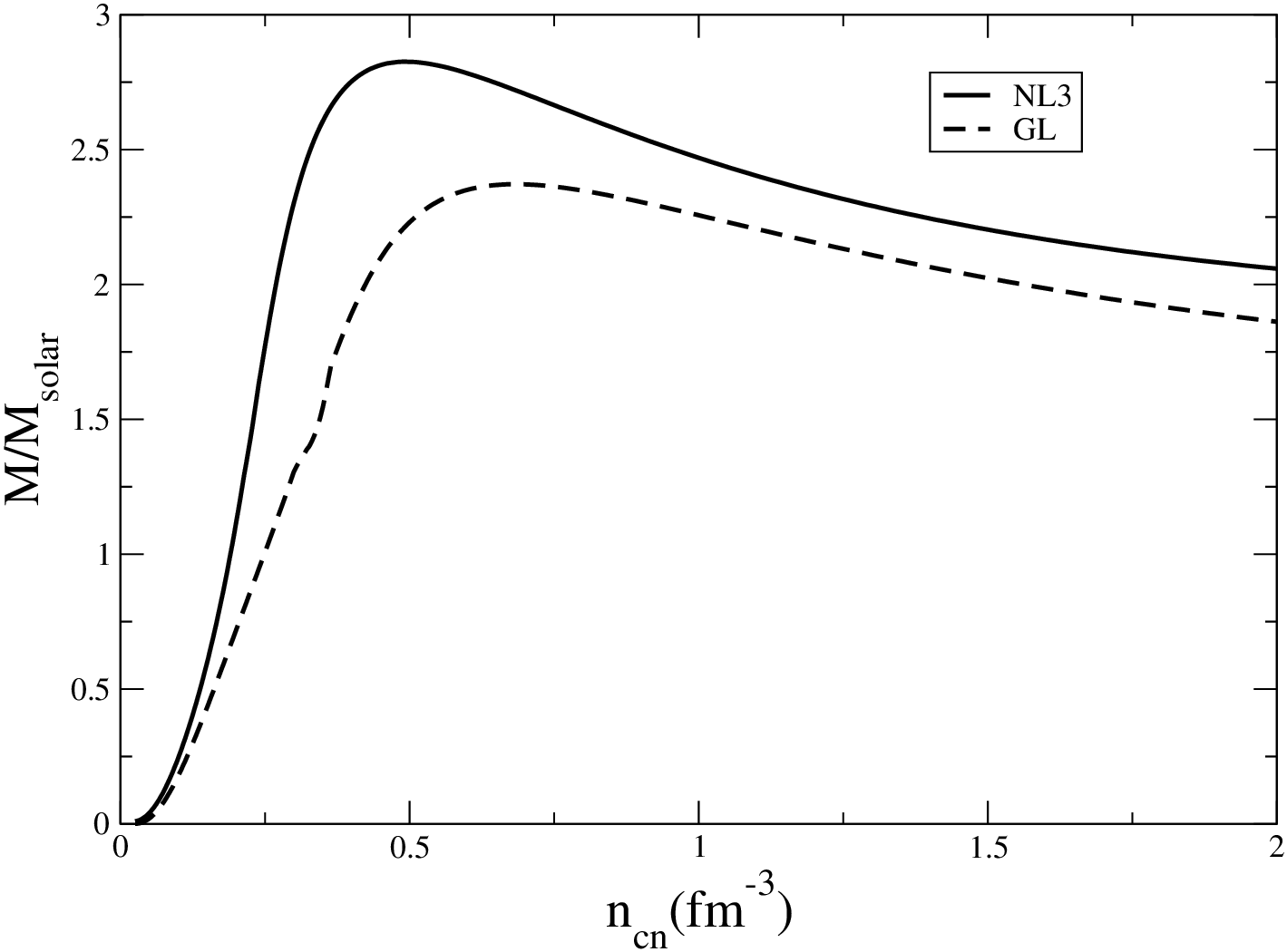}
}}

\vspace{4.0cm}

\noindent{\small{
FIG. 1. Neutron star sequence is plotted with central neutron density. 
The dashed line corresponds to the calculation with the GL parameter set 
whereas the solid line implies that of the NL3 parameter set.}}

\newpage
\vspace{-2cm}

{\centerline{
\epsfxsize=12cm
\epsfysize=14cm
\epsffile{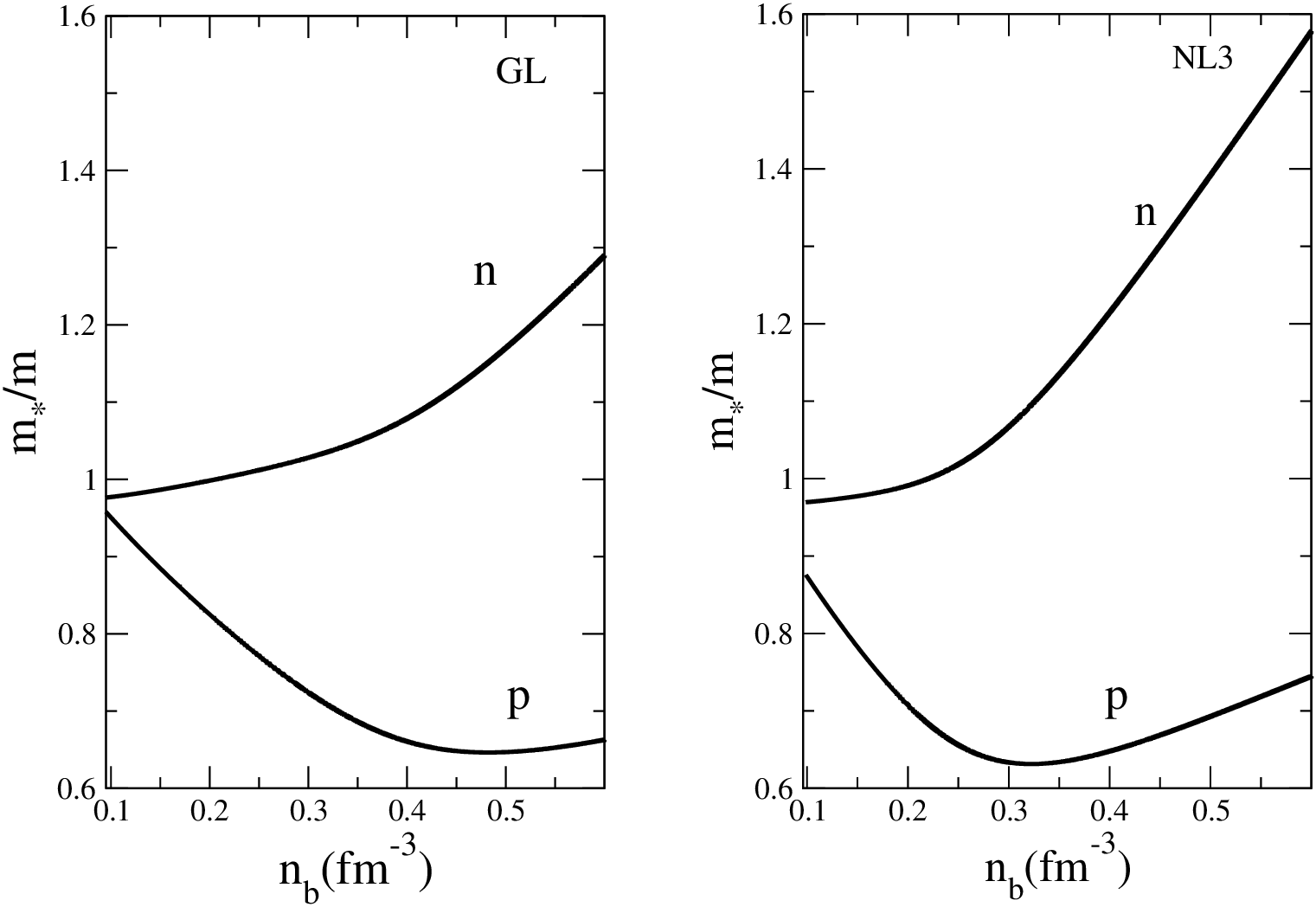}
}}

\vspace{4.0cm}

\noindent{\small{
FIG. 2. Dynamical neutron and proton effective masses are shown as a function 
of baryon density for the GL (left panel) and NL3 (right panel) parameter sets.}}
\newpage
\vspace{-2cm}

\newpage
\vspace{-2cm}

{\centerline{
\epsfxsize=12cm
\epsfysize=14cm
\epsffile{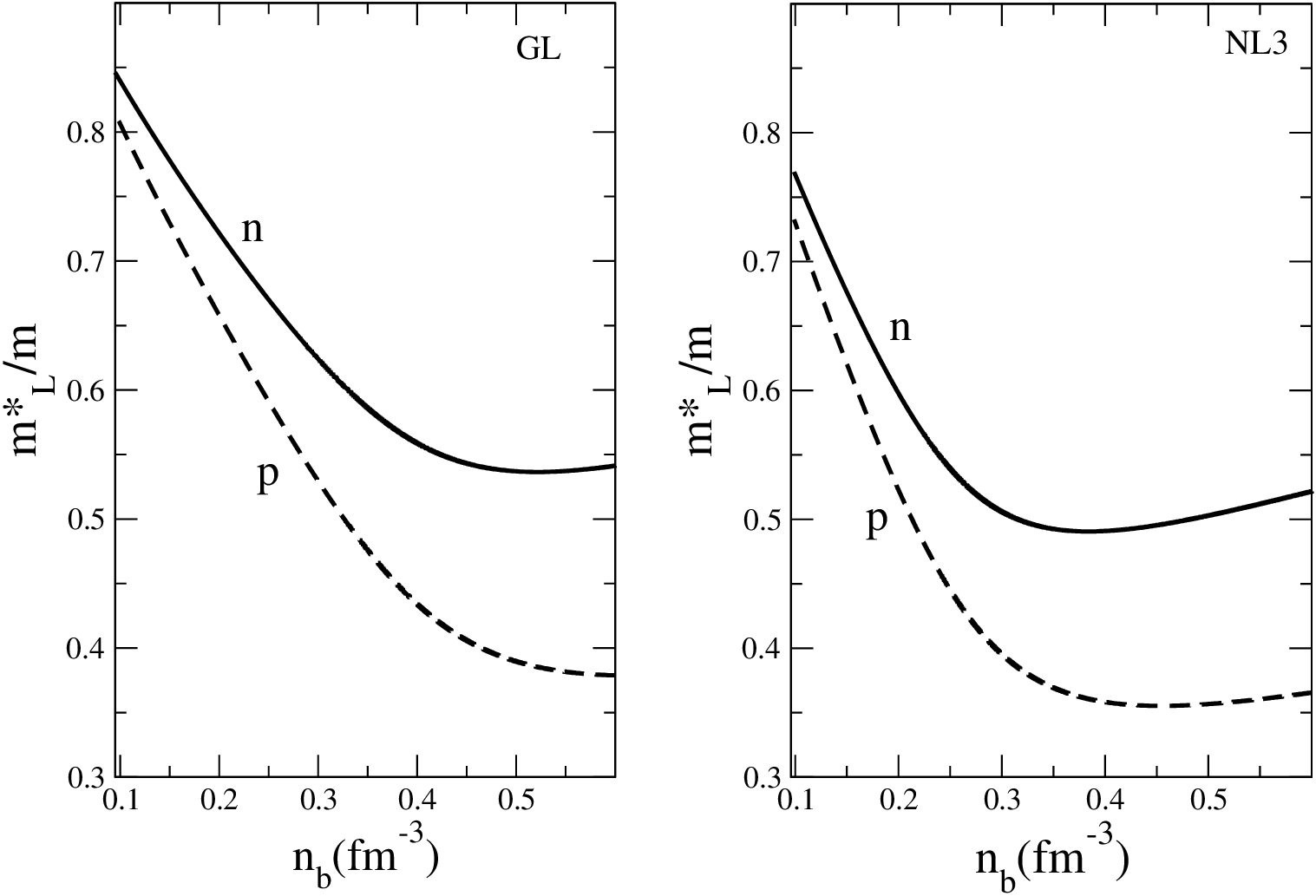}
}}

\vspace{4.0cm}

\noindent{\small{
FIG. 3. Landau effective masses are shown as a function of baryon density for
the GL (left panel) and NL3 (right panel) parameter sets.}}

\newpage
\vspace{-2cm}

{\centerline{
\epsfxsize=12cm
\epsfysize=14cm
\epsffile{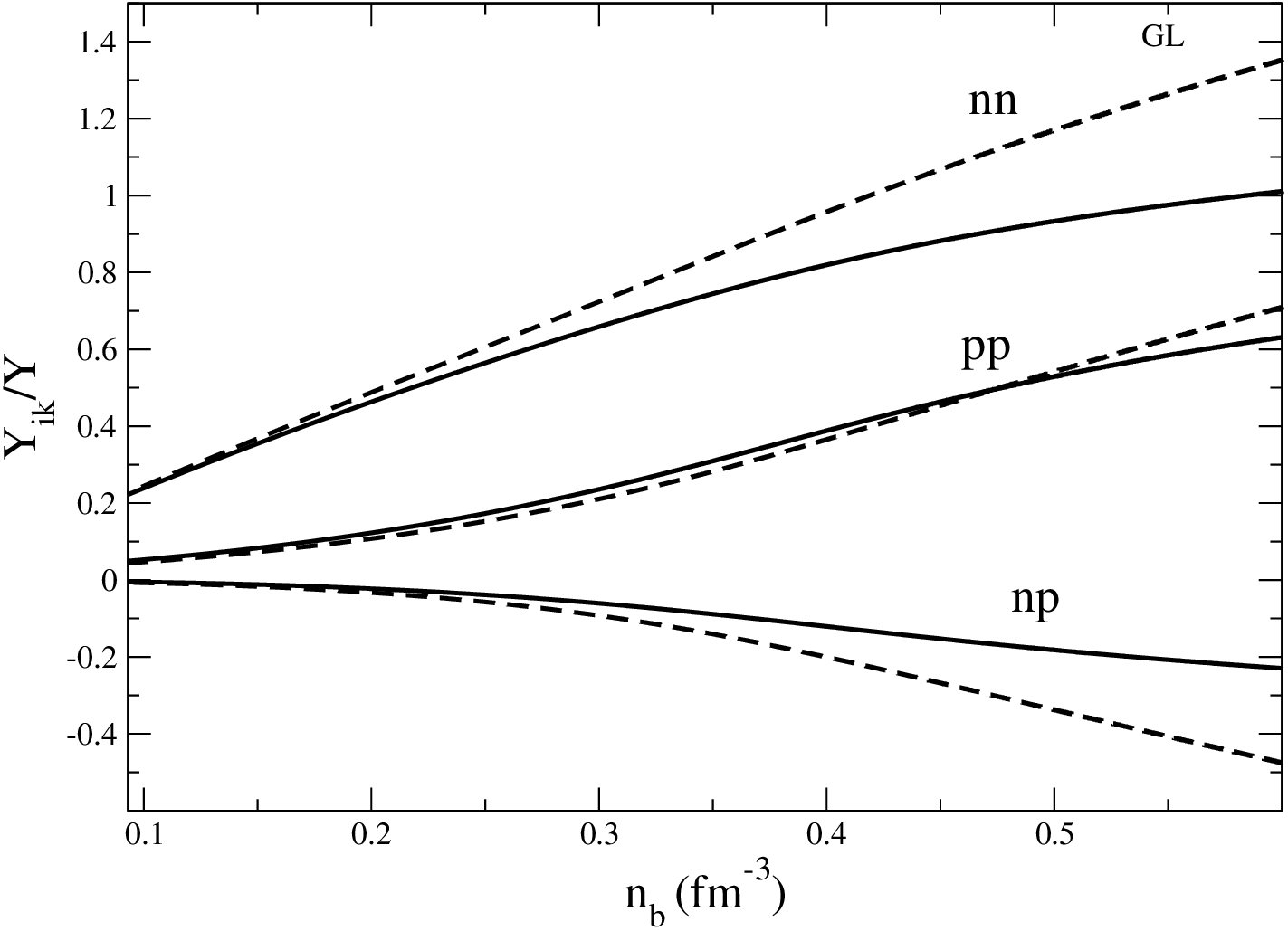}
}}

\vspace{4.0cm}

\noindent{\small{FIG. 4. Normalised entrainment matrix elements ($Y_{ik}/Y$) in 
the zero momentum frame is plotted as a function of baryon density. The 
normalisation factor is taken as $Y=3n_0/{\mu_{n} (3 n_0)}$ \cite{gus09} where 
$n_0$ is the saturation density. Results of this calculation (dashed line) are
compared with those (solid line) of Ref.\cite{gus09}.}}

\newpage
\vspace{-2cm}

{\centerline{
\epsfxsize=12cm
\epsfysize=14cm
\epsffile{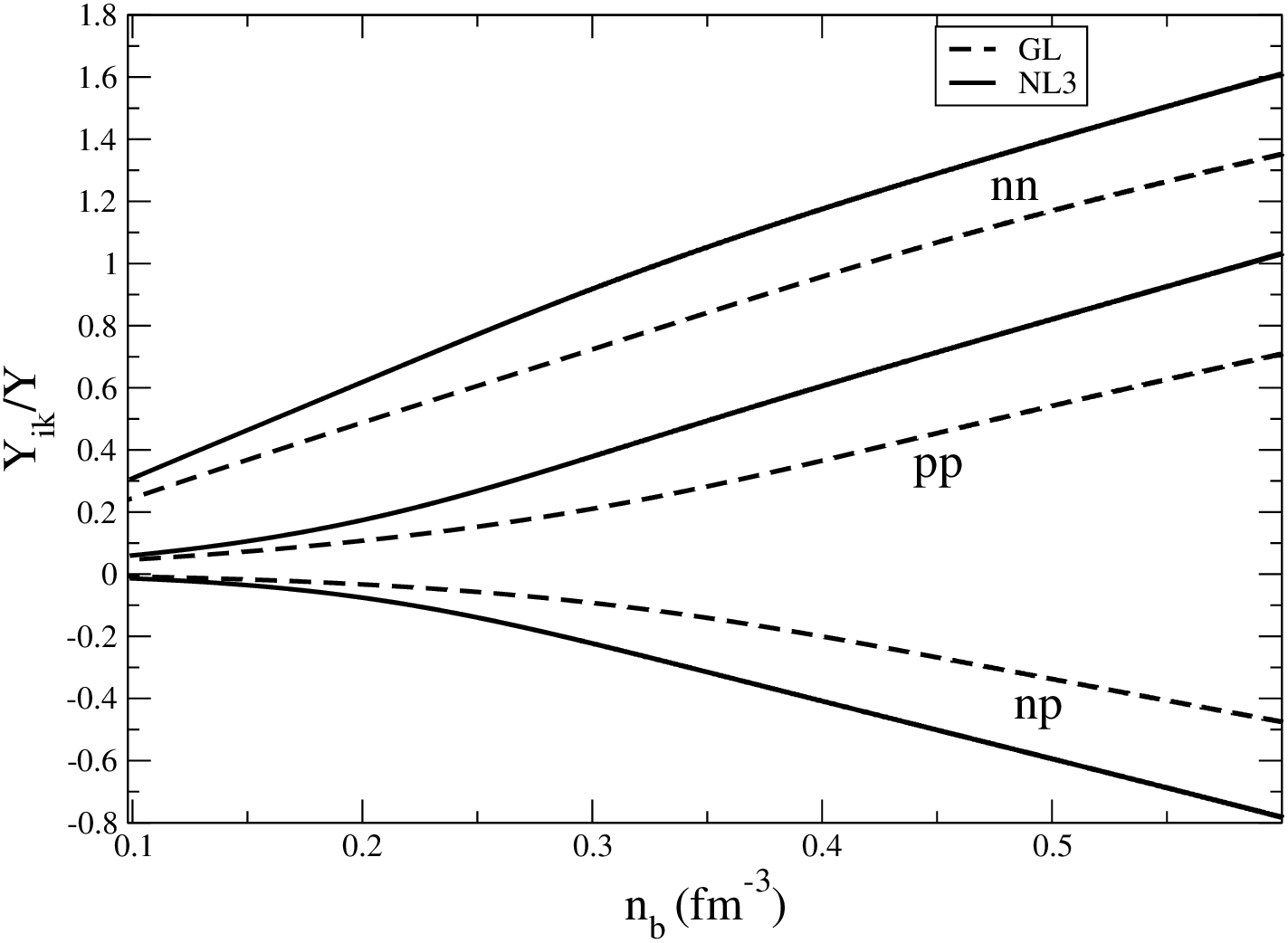}
}}

\vspace{4.0cm}

\noindent{\small{FIG. 5. Normalised entrainment matrix elements ($Y_{ik}/Y$) in 
the zero momentum frame is plotted as a function of baryon density for the GL
and NL3 parameter sets. The 
normalisation factor is taken as $Y=3n_0/{\mu_{n} (3 n_0)}$ \cite{gus09} where 
$n_0$ is the saturation density.}}

\newpage
\vspace{-2cm}

{\centerline{
\epsfxsize=12cm
\epsfysize=14cm
\epsffile{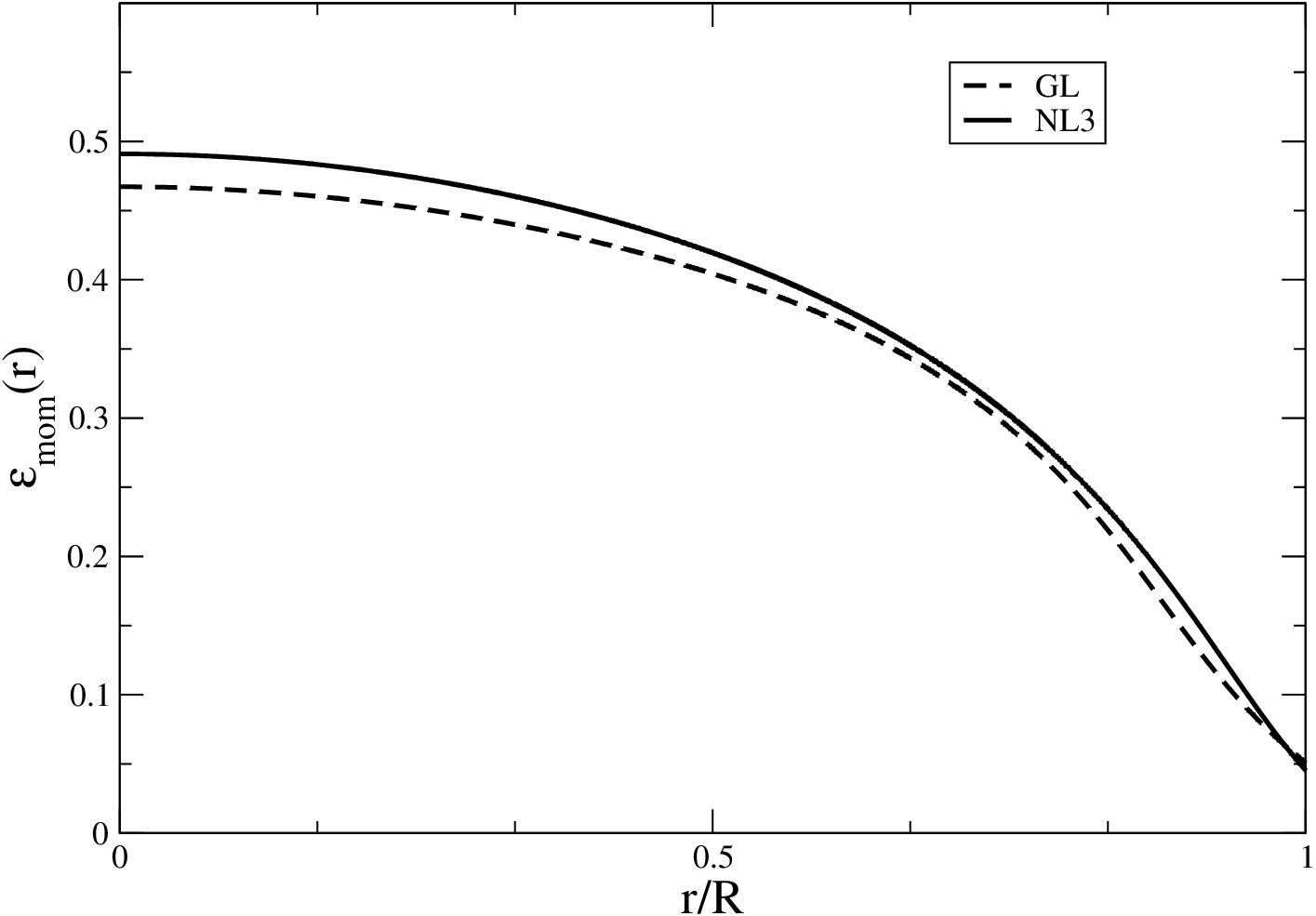}
}}

\vspace{4.0cm}

\noindent{\small{FIG. 6. Entrainment parameter in the RMF model with $\rho$
meson in the zero momentum frame of 
neutrons is plotted as a function of radial distance in a neutron star of
mass $2.37 M_{\odot}$ and radius 11.09 km using the GL parameter set (dashed 
line) and mass $2.82 M_{\odot}$ and radius 13.17 km with the NL3 parameter set
(solid line).}}

\newpage
\vspace{-2cm}

{\centerline{
\epsfxsize=12cm
\epsfysize=14cm
\epsffile{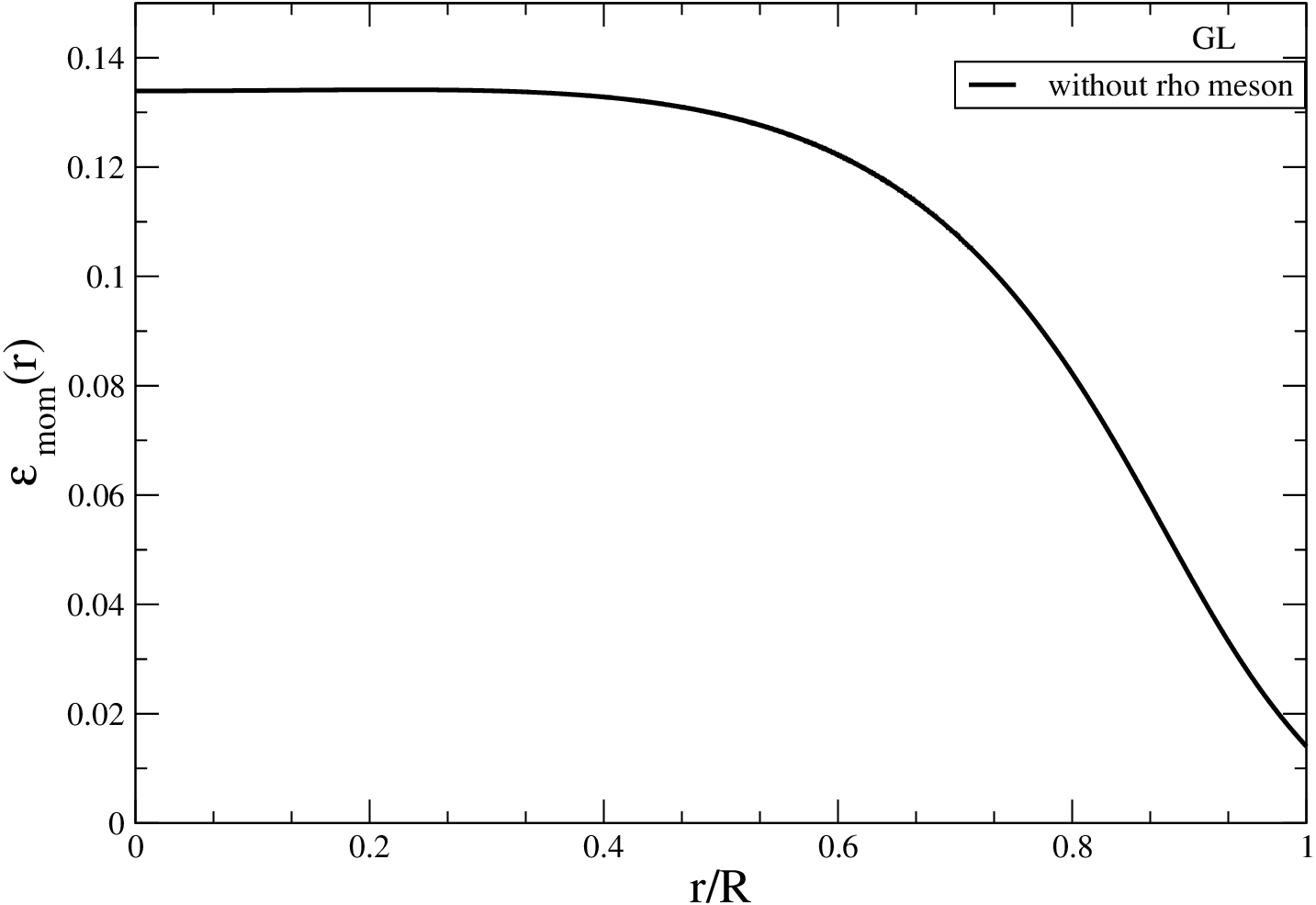}
}}

\vspace{4.0cm}

\noindent{\small{FIG. 7. Entrainment parameter in the RMF model without $\rho$
meson in the zero momentum frame of 
neutrons is plotted as a function of radial distance in a neutron star of
mass 2.33 $M_{\odot}$ and radius 10.96 km.}}

\newpage
\vspace{-2cm}

{\centerline{
\epsfxsize=12cm
\epsfysize=14cm
\epsffile{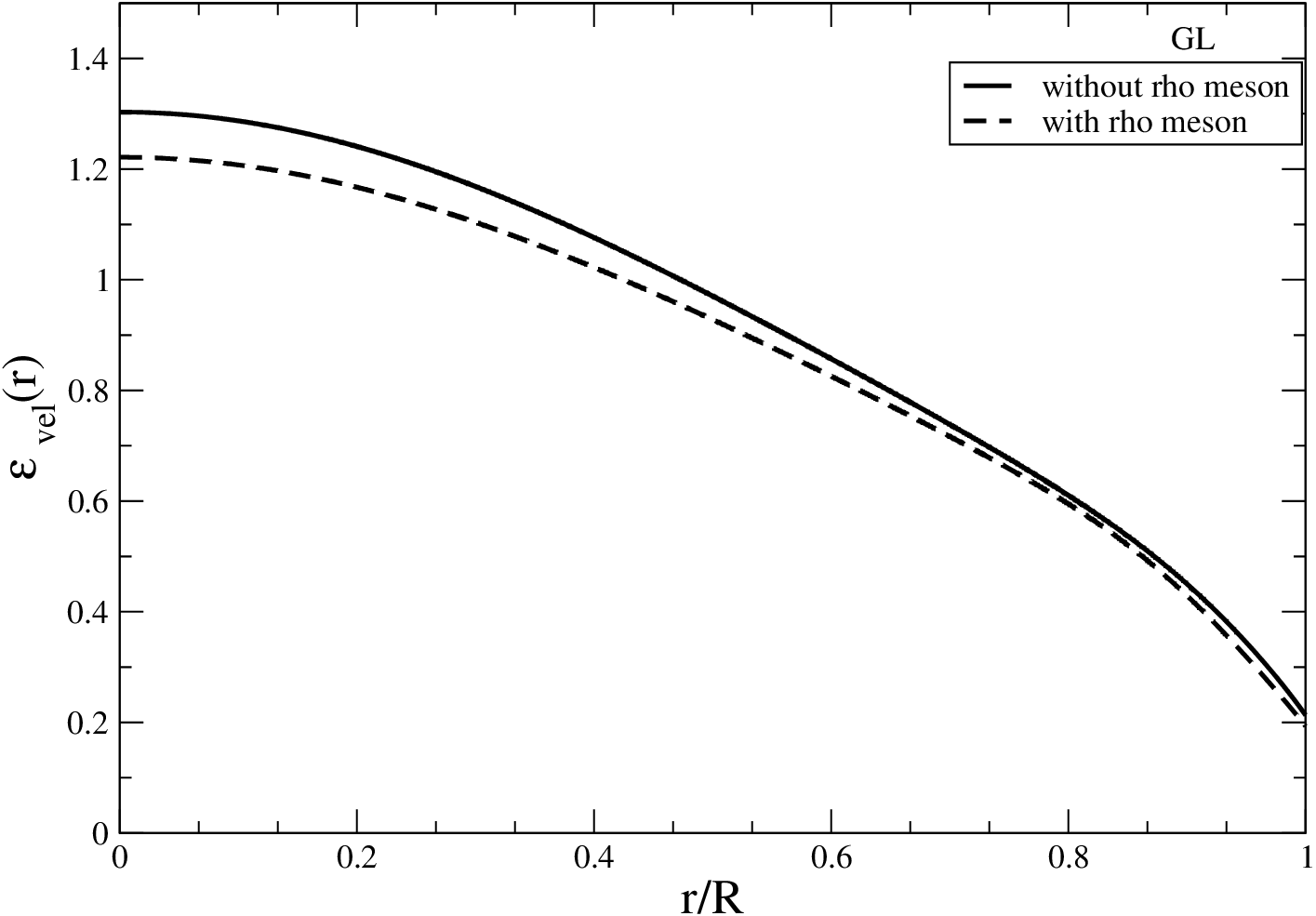}
}}

\vspace{4.0cm}

\noindent{\small{FIG. 8. Entrainment parameter in the zero velocity frame of 
neutrons is plotted as a function of radial distance in neutron stars of masses
2.33 M$_{\odot}$ (solid line) and 2.37 M$_{\odot}$ (dashed line).}}

\newpage
\vspace{-2cm}

{\centerline{
\epsfxsize=12cm
\epsfysize=14cm
\epsffile{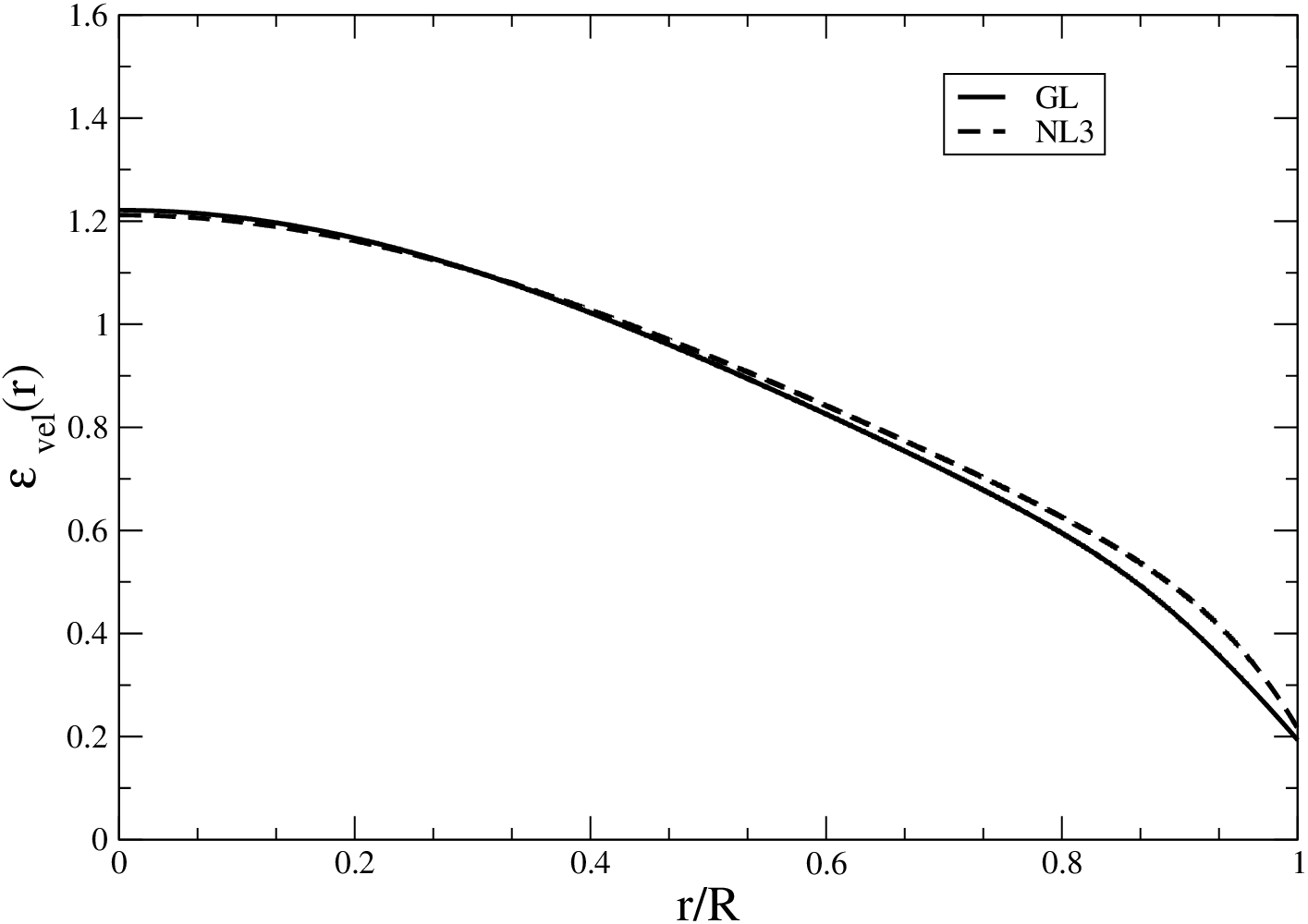}
}}

\vspace{4.0cm}

\noindent{\small{FIG. 9. Entrainment parameter in the zero velocity frame of 
neutrons is plotted as a function of radial distance in neutron stars of masses
2.37 M$_{\odot}$ (solid line) and 2.82 M$_{\odot}$ (dashed line) for the GL and
NL3 parameter sets, respectively.}}
\end{document}